\documentclass[pre,floatfix,superscriptaddress,twocolumn]{revtex4}

\def\av#1{\left\langle#1\right\rangle}
\usepackage{latexsym}
\usepackage{amstext,amsfonts}
\usepackage{epsfig}

\begin{document}

\title{Percolation in networks composed of connectivity and dependency links}

\author{Amir Bashan}
\affiliation{Department of Physics, Bar-Ilan University, Ramat Gan, Israel}
\author{Roni Parshani}
\affiliation{Department of Physics, Bar-Ilan University, Ramat Gan, Israel}
\author{Shlomo Havlin}
\affiliation{Department of Physics, Bar-Ilan University, Ramat Gan, Israel}

\date{\today}

\begin{abstract}
Networks composed from both connectivity and dependency links were found to be
more vulnerable compared to classical networks with only connectivity links. Their percolation transition is
usually of a first order compared to the second order transition found in classical networks.
We analytically analyze the effect of different distributions of dependencies links on the robustness of networks.
For a random Erd$\ddot{o}$s-R$\acute{e}$nyi (ER) network with average degree $k$ that is divided into dependency clusters of size $s$,
 the fraction of nodes that belong to the giant component,
 $P_\infty$, is given by $ P_\infty=p^{s-1}\left[1-\exp{(-kpP_\infty)}  \right]^s $
where $1-p$ is the initial fraction of removed nodes.
Our general result coincides with the known Erd$\ddot{o}$s-R$\acute{e}$nyi
 equation for random networks for $s=1$ and with the result of Parshani et al (PNAS, in press, 2011) for $s=2$.
For networks with Poissonian distribution of dependency links we find that $P_\infty$ is given by
    $P_\infty  =f_{k,p}(P_\infty)e^{(\langle s\rangle -1)(pf_{k,p}(P_\infty)-1)}$ where $f_{k,p}(P_\infty)\equiv1-\exp{(-kpP_\infty)}$ and
$\langle s \rangle$ is the mean value of the size of dependency clusters.
For networks with Gaussian distribution of dependency links we show how the average and width of
the distribution affect the robustness of the networks.
\end{abstract}
\maketitle

\section{Introduction}
Many systems can be efficiently modeled using a network structure where the system entities are the network nodes and the relations between the entities are the network links
\cite{WattsNature, barasci, bararev, PastorXX, mendes, caldarelli1, Reuven_book, caldarelli2, cohena, Vespignani, Albert, Newman2006, NewmanBook, NewmanSIAM}.
However, many systems are also characterized by small sub groups in which the entities belonging to a group strongly depend on each other.
We coin the relation between each two nodes in such a group as dependency links  \cite{Parshani_snc}.
For example consider a financial network: Each company has trading and sales connections with other companies (connectivity links).
These connections enable the companies to interact with others and function together as a global financial market.
In addition companies that belong to the same owner strongly depend on one another (dependency links).
If one company fails the owner might not be able to finance the other companies that will fail too.
Another example is an online social network (Facebook or Twitter): Each individual communicates with his friends (connectivity links),
thus forming a social network through which information and rumors can spread.
However, many individuals will only participate in a social network if other individuals
with common interests also participate in that social network, thereby forming dependency groups.

Previous studies focused on network models containing only a single type of links,
either connectivity links \cite{er1, er1a, bollo, cohena, Callaway, BundeHavlinBook}
or dependency links \cite{Motter2002, Motter, Sachtjen, Moreno, Watts2002}.
A network model containing both connectivity and dependency links was first introduced for two interdependent networks
\cite{Buldyrev,Parshani_PRL}.

\begin{figure}
\begin{center}
\epsfig{file=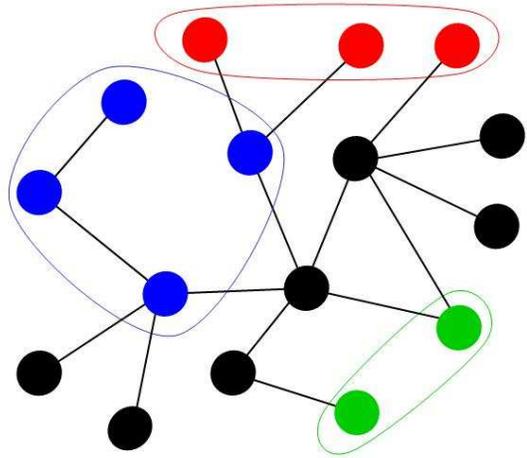, width=9cm}
\end{center}
\caption{(Colors online) Connectivity network with dependency clusters. The edges represent
connectivity relations while the (blue, red and pink) groups surrounded by curves represent dependency
relations between all the nodes of the same group (color). The dependency relations
can be between very $"far"$ nodes, in the connectivity network. In the general case,
the sizes of each dependency clusters follow a given distribution.  }
\label{Illustration}
\end{figure}

 A recent  paper \cite{Parshani_snc} introduced for the first time a single network model containing both connectivity and dependency links.
In this network model the initial failure of nodes may trigger an iterative process of cascading failures that has a devastating effect on the network stability. The cascading failures are a result of the synergy between two different effects: a) A percolation process governed by connectivity links. b) The failure of an entire dependency groups due to a failure of one member within the group.
 For a high density of dependency links the network disintegrates in a form of a first order phase transition while for a low density of dependency links the network disintegrates in a second order transition.

However the combined model presented in \cite{Parshani_snc} was based on an unrealistic assumption that all dependency groups are of size 2, i.e. only pair of nodes depend on each other. In reality, as the examples above suggest, groups of several elements may depend on each other.

In this paper we analyze both analytically and numerically the general case of a network with different sizes of dependency clusters, as illustrated in Fig. \ref{Illustration}. We study networks with three different types of dependency groups:
a) Fixed size, $s$, dependency groups. b) Normally distributed sizes of dependency groups. c) Poisson distributed sizes of dependency groups.
We find that for random networks with an average degree $k$ which are divided into dependency groups (clusters) of size $s$,
the fraction of nodes belonging to the giant component, $P_\infty$, is given by $ P_\infty=p^{s-1}\left[1-\exp{(-kpP_\infty)}  \right]^s $,
where $1-p$ is the initial fraction of removed nodes. The critical threshold, $p_c$, below which the network
collapses $(P_\infty=0)$ is given by Eq. (\ref{p_c}). Our result for $s=1$ (a node depends only on itself) coincides with the
known Erd$\ddot{o}$s-R$\acute{e}$nyi equation, $ P_\infty=1-\exp{(-kpP_\infty)}$, for a network without dependency relations \cite{er1, er1a, bollo}.
We also show that for $s\geq2$ a process of cascading failures occurs and the percolation transition is of first order.

For normally distributed dependency groups with an average size $\langle s \rangle$ and width $\sigma$, we find that
the system becomes more stable (smaller $p_c$) for a broader size distribution (Fig.\ref{PcVsSigma}). When $\sigma\rightarrow 0$ the results are the same as the case of fixed size dependency groups with $\langle s \rangle=s$ (Fig.\ref{Pcvss}).
We also analyze both analytically and numerically the case of a Poisson distribution of
dependency cluster sizes and obtain analytical equations for both $P_\infty(\av{s})$ and $p_c$
(Eqs. (\ref{P_inf_trans_Bino}),(\ref{P_II}) and (\ref{derivatives_Bino})).

\section{General Formalism}
When nodes fail in a network containing both connectivity links and dependency clusters, two different processes occur.
(i) Connectivity links connected to these nodes fail, causing other nodes to disconnect from the network (percolation step).
(ii) A failing node cause the failure of all the other nodes of its dependency cluster,
 even though they are still connected via connectivity links (dependency step).
Thus, a node that fails in the percolation step leads to the failure of its entire dependency cluster, which in turn leads to a new
percolation step, which further leads to a dependency step and so on. Once the cascade process is triggered it will only stop if nodes that fail in one step do not cause additional failure in the next step.

We start by presenting the formalism describing the iterative process of cascading failures.
On each step we apply the two processes - a percolation process followed by the removal of relevant dependency groups.
Before each percolation stage the accumulated cascades are described as equivalent to a single random removal, $1-\psi^p_n$.
Similarly, before each dependency stage the accumulated cascades are equivalent to a single random removal, $1-\psi^D_n$.
When applying the percolation process at stage $n$ on a network of size $\psi^p_n$
 the remaining giant component consists  of a fraction $g_p(\psi^p_n)$ which is a fraction
 $\phi^p_n=\psi^p_n g_p(\psi^p_n)$  from the original network.
Similarly, applying the dependency process at stage $n$ on a network of size $\psi^D_n$
results in a remaining functional nodes consisting of a fraction $g_D(\psi^D_n)$ which is a fraction
$\phi^D_n=\psi^D_n g_D(\psi^D_n)$ from the original network.

The iterative process is initiated by the removal of a fraction $1-p$ of the network nodes.
The remaining part of the network is $\psi^p_1 \equiv p$. This initial removal will cause additional
nodes to disconnect from the giant cluster due to the percolation process. The fraction of nodes
that remain functional after the percolation process is $\phi^p_1=\psi^p_1 g_P(\psi^p_1) $.
Before the dependency step we describe the accumulated cascades of the previous steps.
The fractions of nodes that fail due to the initial removal and due to first percolation step
are  $1-\psi^p_1$ and $\psi^p_1-\phi^p_1$, respectively,
and the accumulated cascades are equivalent to a single random removal of
$1-\phi^p_1=(1-\psi^p_1)+(\psi^p_1-\phi^p_1)$.
Thus, we denote the remaining functional part
 before the dependency step as $\psi^D_1\equiv \phi^p_1$.
Each node from the non functional part will cause all the other nodes of its dependency
 cluster to also fail (dependency process). The remaining functional part of the network after
 the dependency step is $\phi^D_1=\psi^D_1 g_D(\psi^D_1)$.

Let us now calculate  the accumulated failure up to this step. The sum of the
previous steps, the initial removal of (1-p), the removal due to the percolation step,
 $(\psi^p_1-\phi^p_1)$ and the removal due to the dependency step $(\psi^D_1-\phi^D_1)$, is equivalent
to a single random removal of $(1-pg^D(\psi^D_1))$ from the original network (\cite{Parshani_snc}).
After such removal the remaining part of the network before the second percolation step is
$\psi^p_2 \equiv pg^D(\psi^D_1)$, and the size of giant cluster is then
$\phi^p_2=\psi^p_2 g_p(\psi^p_2)$.

Following this approach we can construct the sequences $\psi^p_n$ and $\psi^D_n$ of the remaining fraction of nodes
and the sequences $\phi^p_n$ and $\phi^D_n$ of functional nodes, at each stage of the cascade of failures.
The general form is given by:

\begin{eqnarray}
    \begin{array}{cl}
        &\psi^p_1\equiv p , \phi^p_1=\psi^p_1 g_p(\psi^p_1),\\
        &\psi^D_1=g_p(\psi^p_1)p, \phi^D_1=\psi^D_1 g_D(\psi^D_1),\\
        &\psi^p_2= g_D(\psi^D_1)p, \phi^p_2= \psi^p_2 g_p(\psi^p_2)\\
        &\vdots\\
        &\psi^p_n=g_D(\psi^D_{n-1})p, \phi^p_n=\psi^p_n g_p(\psi^p_n),\\
        &\psi^D_{n}=g_p(\psi^p_{n})p, \phi^D_{n}=\psi^D_{n} g_D(\psi^D_{n})
    \end{array}
    \label{cascade}
\end{eqnarray}

\noindent To determine the state of the system at the end of the
cascade process we look at $\psi^p_{m}$ and $\psi^D_{m}$ at the limit
of $m \to \infty$.  This limit must satisfy the equation
$\psi^p_{m}$=$\psi^p_{m+1}$ (or $\psi^D_{m}$=$\psi^D_{m+1}$) since
eventually the clusters stop fragmenting and the fractions of
randomly removed nodes at step $m$ and $m+1$ are equal. Thus,
at steady state the system satisfies the set of two equations
\begin{eqnarray}
\nonumber \psi^p_\infty&=&g_D(\psi^D_{\infty})p, \\
          \psi^D_{\infty}&=&g_p(\psi^p_{\infty})p.
\end{eqnarray}

  Denoting
$x\equiv \psi^D_{\infty}$ and $y\equiv \psi^p_{\infty}$ we arrive to a system of two
equations with two unknowns: $x=pg_p(y)$  $y=pg_D(x)$
which can be reduced to:

\begin{equation}
\label{equation_x}
x=pg_p(pg_D(x)).
\end{equation}

\noindent Solving equation (\ref{equation_x}) we obtain the size of the
network at the end of a cascade initiated by random removal of $1-p$ of the nodes.

Next, we calculate explicitly $g_D(T)$ and $g_p(T)$.
In the general case, each node belongs to a dependency group of size $s$ with a
probability $q(s)$ so that the number of groups of size $s$ is equal to $q(s)N/s$.
Since after random removal of $1-T$ of the nodes each group of size $s$ remains functional
with a probability $T^s$, the total number of nodes that remain functional
is given by $\sum_{s=1}^{\infty}q(s)NT^s$. Thus, we define the function $g_D(T)$ as
the fraction of nodes that remain functional out of the $TN$ nodes that were not removed,
\begin{equation}\label{gD}
    g_D(T)\equiv\sum_{s=1}^{\infty}q(s)T^{s-1}.
\end{equation}

Analogous to $g_D(T)$, $g_p(T)$ is defined as the fraction of nodes belonging to the giant cluster
of the connectivity network after random removal of $1-T$ of the nodes. The percolation process
can be solved analytically by using the apparatus
of generating function. As in Refs.~\cite{Newman,Shao,Shao09} we will introduce the generating
function of the degree distributions $G_{0}(\xi)=\sum_k P(k) \xi^k$.
Analogously, we will introduce the generating function of the underlining
branching processes, $G_{1}(\xi)=G'_{0}(\xi)/ G'_{0}(1)$.
Random removal of fraction $1-T$ of nodes will change the degree
distribution of the remaining nodes, so the generating function of the
new distribution is equal to the generating function of the original
distribution with the argument equal to $1-T(1-\xi)$
\cite{Newman}. The fraction of nodes that belong to the giant
component after the removal of $1-T$ nodes is \cite{Shao,Shao09}:
\begin{equation}
g_p(T)=1-G_{0}[1-T(1-u)],
\label{e:g}
\end{equation}
\noindent where $u=u(T)$ satisfies the self-consistency relation
\begin{equation}
u=G_{1}[1-T(1-u)].
\label{e:u}
\end{equation}

\section{ER Networks}
The formalism presented in Sect. II is general for a random network having any degree distribution.
In the case of an ER network, whose degrees are Poisson-distributed
\cite{er1,er1a,bollo}, the problem can be solved explicitly. Suppose that
the average degree of the network is $k$. Then, $G_{1}(\xi)=G_{0}(\xi)=\exp[k(\xi-1)]$.
Thus, $g_p(x)=1-u$
and therefore Eq. (\ref{equation_x}) becomes
\begin{equation}\label{x_eq}
    x=p[1-u],
\end{equation}

\begin{figure}
\begin{center}
\epsfig{file=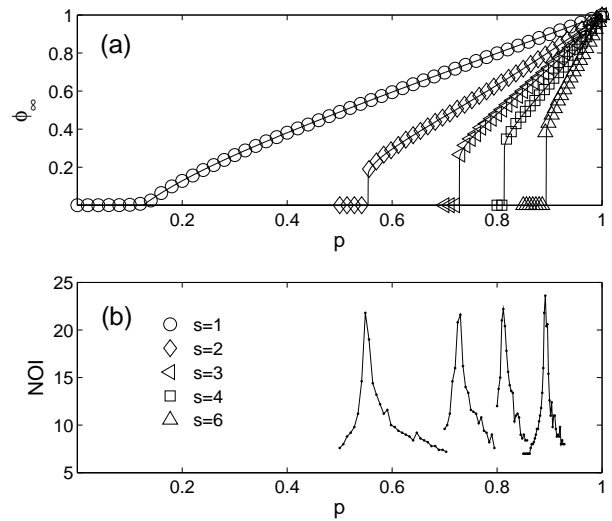,width=9cm}
\end{center}
\caption{(a) The size of the giant cluster, $\phi_\infty\equiv P_\infty \cdot p$, vs.
$p$, the fraction of nodes that remain after random removal, for ER networks ($k=8$)
 for different fixed sizes of dependency clusters, $s$. The symbols
represent simulation results of systems of $50,000$ nodes and the solid lines
show the theoretical predictions. For the case of
$s=1$ there are no dependency clusters and the regular percolation process
leads to second order phase transition. For $s\geq2$, a first order phase transition
characterize the percolation process represented by discontinuity of $P_\infty$ at $p_c$. Both the regular and the new first order percolation
obey Eq. (\ref{trans_P_inf_fixed_size}).
(b) The number of iterative failures (NOI) sharply increases when approaching the critical
threshold $p_c$ for the first order transitions, and thus represent a useful method for identifying
accurately the value of $p_c$ \cite{Parshani_snc}. In the figure, each curve is maximal as its
related curve in panel (a) approaches from both sides the critical threshold.}
\label{PinfNOIvspFixedSize}
\end{figure}

\noindent where $u$  is defined according to (\ref{e:u}), by
\begin{equation}\label{trans_eq_u}
    u=\exp(-kpg_D(x)(1-u)).
\end{equation}

Using the definition of $g_D(x)$, Eq. (\ref{gD}), together with Eq. (\ref{trans_eq_u})
we get the general solution for the steady state of the network at the end of the cascade failure process

\begin{equation}\label{u_trans}
    u=e^{-k\sum_{s=1}^\infty q(s)p^s(1-u)^s}.
\end{equation}

In order to present $u$, obtained from Eq. (\ref{u_trans}), in terms of $P_\infty$ recall that
at the steady state the size of the giant cluster $\phi_\infty\equiv\phi^p_n=\phi^D_n$, and according to (\ref{cascade})
\begin{equation}\label{steady_state}
    \phi_\infty=xg_D(x)=x\sum_{s=1}^\infty q(s)x^{s-1}=\sum_{s=1}^\infty q(s)x^s.
\end{equation}

 Since $P_\infty \equiv \phi_\infty / p$, we get the relation,
\begin{equation}\label{P_inf_u}
  P_\infty \cdot p=\sum_{s=1}^\infty q(s)p^s(1-u)^s
\end{equation}
and using (\ref{u_trans}), a simple equation for $P_\infty$ is obtained
\begin{equation}\label{P_inf_ln_u}
    P_\infty=-\frac{\ln u}{kp},
\end{equation}
where $u$ is the solution of Eq. (\ref{u_trans}).

Up to this point, we obtained the size of the network at each step
of the cascade process, Eq. (\ref{cascade}), and in particular, its size, $P_\infty$, Eqs. (\ref{u_trans})
 and (\ref{P_inf_ln_u}), at the end of the cascade,
for the general case of a given distribution $q(s)$ of sizes of dependency clusters.

\section{Fixed Size of Dependency Clusters}
Using the general solution described above, we analyze the case of a
fixed size, $s$, of dependency clusters. In particular, we find the
size of the giant component and the critical
fraction of the network, $1-p_c$, that, if removed, leads to complete fragmentation
of the network. In this case, $g_D$, given in (\ref{gD}), becomes
$g_D(T)=T^{s-1}$ and Eqs. (\ref{u_trans}) and (\ref{P_inf_u}) become respectively:\\
\begin{eqnarray}
\label{trans_u_fixed}
  u &=& e^{-kp^s(1-u)^s} \\
  P_\infty &=& p^{s-1}(1-u)^s
\end{eqnarray}

\noindent which can be combined into a single equation:
\begin{equation}\label{trans_P_inf_fixed_size}
    P_\infty=p^{s-1}\left(1-e^{-kpP_\infty}  \right)^s.
\end{equation}

Eq. (\ref{trans_P_inf_fixed_size}) coincides for $s=1$ (a node depends only on itself) with the
known Erd$\ddot{o}$s-R$\acute{e}$nyi equation \cite{er1, er1a, bollo}, $ P_\infty=1-\exp{(-kpP_\infty)}$, for a network without dependency relations. Moreover, for $s=2$, Eq. (\ref{trans_P_inf_fixed_size}) yields the result
obtained in \cite{Parshani_snc} for the case of dependency pairs.

Fig. \ref{PinfNOIvspFixedSize} shows the size of the giant cluster, $P_\infty$, versus the fraction of
nodes, $p$, remaining after an initial random removal of $1-p$, for the case of ER network with fixed size of dependency clusters $s$. The case of $s=1$, each node depend only on itself, is the regular second order
percolation transition. For any $s\geq2$, a first order phase transition
characterizes the percolation process. Both the regular and the new first order percolation
obey Eq. (\ref{trans_P_inf_fixed_size}).

Finding the transition point via simulations is always a difficult task that requires high precision. In the case of $s\geq2$ where first order transition occurs we are able to calculate the transition point with good precision by identifying the special behavior characterizing the number of iterations (NOI) in the cascading process \cite{Parshani_snc}. This number sharply drops as the distance from the transition point is increased. Thus, plotting the NOI as a function of \textit{p}, provides a useful and precise method for identifying the transition point $p_c$ in the first order region. Fig. \ref{PinfNOIvspFixedSize}b presents NOI of the simulation results of Fig. \ref{PinfNOIvspFixedSize}a.
The transition point, $p_c$, can easily be identified by the sharp peak characterizing the percolation threshold.
The results shown in Fig. 2b are in excellent agreement with theory.

Next, we find analytically the percolation threshold, $p_c$, for the case of a fixed size of dependency clusters.
Eq. (\ref{trans_u_fixed}), which is the condition for a steady state, have a trivial solution at $u=1$, corresponds, by (\ref{P_inf_ln_u}),
 to a complete fragmentation of the network. For large $p$ there is another solution of $0<u<1$, corresponding to a finite fraction of the network.
Therefore, the critical case corresponds to satisfying both the tangential condition for Eq. (\ref{trans_u_fixed}),
\begin{equation}\label{derivative_u_fixed}
    1=u\left[ kp^{s}s(1-u)^{s-1} \right],
\end{equation}
as well as Eq. (\ref{trans_u_fixed}). Thus, combining Eqs. (\ref{derivative_u_fixed})
and (\ref{trans_u_fixed}) we get a closed-form expression for the critical value, $u_c$,

\begin{equation}\label{critical_u_fixed}
        u_{c}=\exp{\left(\frac{u_{c}-1}{su_{c}}\right)}.
\end{equation}
Once $u_c$ is known, we obtain $p_c$ by substituting it into Eq. (\ref{derivative_u_fixed})

\begin{equation}\label{p_c}
    p_{c}=\left[ksu_{c}(1-u_{c})^{s-1} \right]^{-1/s}.
\end{equation}

\begin{figure}[t]
\begin{center}
\epsfig{file=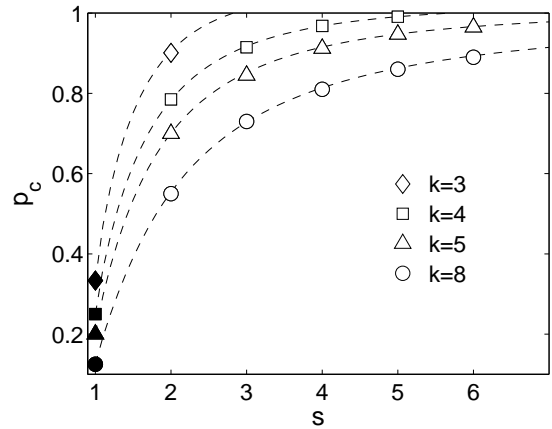,width=8cm}
\end{center}
\caption{Theory (Eq. (\ref{p_c}), dashed lines) and simulation results (symbols) are compared for the values
of $p_c$ for ER networks with different average degree $k$ and different fixed sizes of dependency
clusters, $s$. The full symbols, for $s=1$ represent the known  Erd$\ddot{o}$s-R$\acute{e}$nyi
second-order phase transition threshold for a network without dependency relations, while open symbols represent
first-order phase transition thresholds.}
\label{Pcvss}
\end{figure}

\begin{figure}[t]
\begin{center}
\epsfig{file=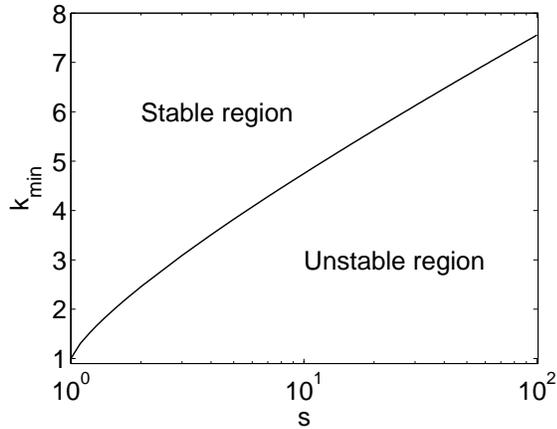,width=8cm}
\end{center}
\caption{The minimum averaged degree, $k_{min}$, as a function of the (fixed) size of dependency clusters, $s$.
The solid line shows the theoretical results, obtained from Eqs. (\ref{critical_u_fixed})
and (\ref{k_min}). The region of $s$ and $k$ values below the line represents an unstable network that will collapse after any single node failure. For $k$ and $s$ values above the line, the network is stable and there exists $p_c<1$.}
\label{KminVsS}
\end{figure}

For $s=1$ we obtain the known result $p_c=1/k$ of Erd$\ddot{o}$s-R$\acute{e}$nyi \cite{er1, er1a, bollo}. Substituting
$s=2$ in Eqs. (\ref{critical_u_fixed}) and (\ref{p_c}) one obtains ${p_c}^2=1/[2ku_c(1-u_c)]$,
which coincides with the exact result found in ~\cite{Parshani_snc}.
 In Fig. \ref{Pcvss} we plot the values of $p_c$ as a function of $s$ for several $k$ values. Note  the dramatic effect of the dependencies on the vulnerability of the system. Even for high values of $k$, $p_c$ approach rapidly  to $1$ even for relatively small s values.


Next, we show that the difference between the continuous second order percolation transition and the first order transition is
characterized not only by the abrupt jump in the size of the giant cluster at the critical
point, $p_c$, but also by a difference in the scaling behavior of the giant
component, $\phi_\infty$, near $p_c$. The scaling near $p_c$ is defined by the exponent $\beta$,
\begin{equation}\label{beta}
    \phi_{\infty}(p)-\phi_{\infty}(p_c)\sim(p-p_c)^{\beta}.
\end{equation}

Eq. (\ref{trans_P_inf_fixed_size}) can be written in terms of $\phi_\infty(\equiv P_\infty \cdot p)$,
the size of the giant component,
 \begin{equation}\label{Phi_inf}
    \phi_\infty=p^s(1-e^{-k\phi_\infty})^s.
 \end{equation}
For the case of $s=1$ (ER) $\phi_{\infty}(p)-\phi_{\infty}(p_c)$ changes (as well known \cite{BundeHavlinBook}) linearly
with $p-p_c$ and $\beta=1$.
For $s\geq2$, we calculate the behavior close to (and above) the critical point,
\begin{eqnarray*}
  p &\equiv& p_c+\epsilon \nonumber\\
  \phi_\infty &\equiv& \phi^c_\infty + \delta \nonumber
\end{eqnarray*}
when $\delta , \epsilon \rightarrow 0$ and $\phi^c_\infty \equiv \phi_{\infty}(p_c)$.
For this case, Eq. (\ref{Phi_inf}) can be written as
 \begin{equation}\label{p_phi_inf}
    p_c+\epsilon=\frac{{(\phi^c_\infty + \delta)}^{1/s}}{1-e^{-k(\phi^c_\infty + \delta)}}
    =A \left[ 1+C_1\delta +C_2\delta^2+\ldots  \right]
 \end{equation}
 where $A\equiv \frac{(\phi^c_\infty)^{1/s}}{1-e^{-k\phi^c_\infty}}= p_c$ and the linear
 coefficient is given by $C_1\equiv \left( \frac{1}{s \phi^c_\infty} - \frac{k}{e^{k\phi^c_\infty}-1}  \right)$.
However, using Eqs. (\ref{P_inf_ln_u}) and (\ref{critical_u_fixed}) we obtain that $C_1=0$,
so near the critical point $\epsilon\sim\delta^2$  and $\delta\sim\epsilon^{1/2}$.
Thus, the scaling behavior of the giant component near the first order transition, Eq. (\ref{beta}),
is characterized by the critical exponent $\beta=1/2$.

For a fixed $s$, when $k$ is smaller than a critical number $k_{min}(s)$, $p_c \geq 1$, meaning that for  $k<k_{min}(s)$, the network will collapse for any finite number of nodes failure.
From Eq. (\ref{p_c}) we get the minimum of $k$ as a function of $s$,

\begin{equation}\label{k_min}
    k_{min}(s)=\left[ su_c(1-u_c)^{s-1} \right]^{-1}.
\end{equation}

Fig. \ref{KminVsS} shows the minimum averaged degree, $k_{min}$, as a function of the size of dependency clusters, $s$.

\begin{figure}[!t]
\begin{center}
\begin{tabular}{cc}
\epsfig{file=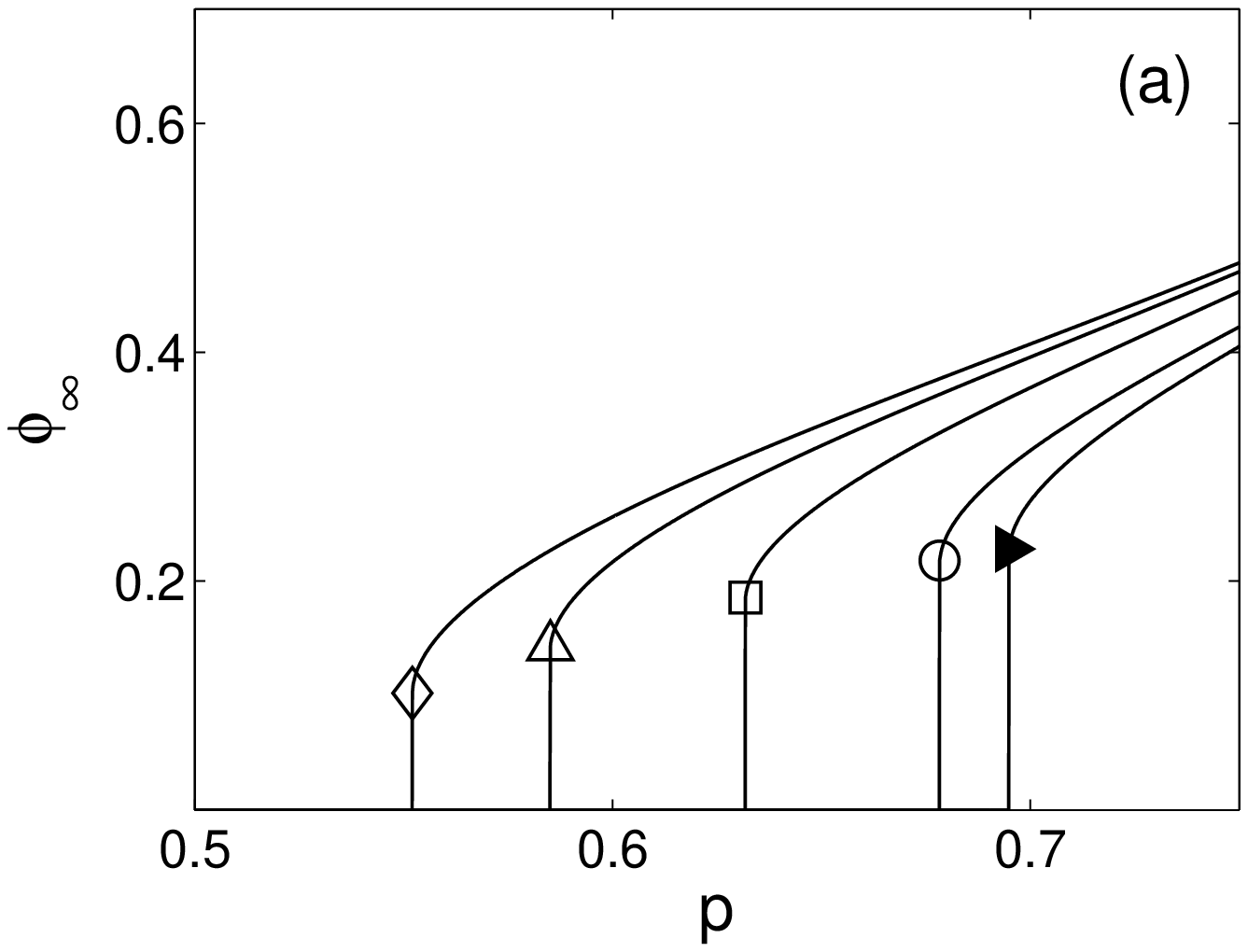,width=8.2cm}\\
\epsfig{file=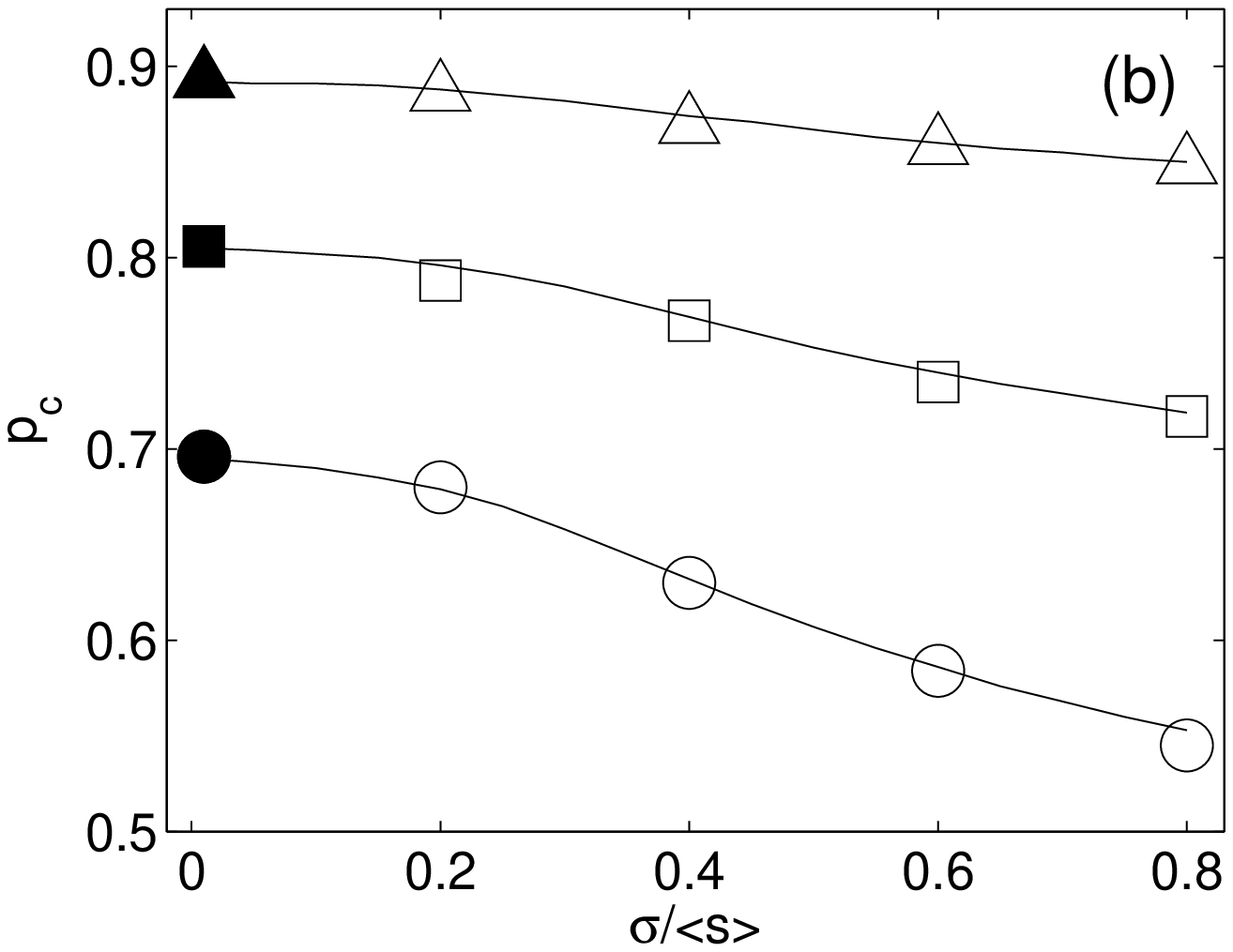,width=8cm}
\end{tabular}
\end{center}
\caption{(a) The size of the giant cluster, $\phi_\infty\equiv P_\infty \cdot p$, vs. p (solid lines)
and $\phi_\infty(p_c)$ (symbols), for ER networks with
 average degree $k=15$ and dependency clusters normally distributed around averaged size
 $\langle s \rangle=4$. The different curves represent different standard deviations, $\sigma$ ($\sigma=0$
 (full triangle right), $\sigma=0.8$ (circle), $\sigma=1.6$ (square), $\sigma=2.4$ (triangle up)
 and $\sigma=3.2$ (diamond)).
(b) Theory (solid lines) and simulations (symbols) values of $p_c$ for ER networks with
 average degree $k=15$ and dependency
 clusters normally distributed around averaged size $\langle s \rangle$, versus the width
  of the distribution, $\sigma$. The three curves represent different values of <s>: $\langle s \rangle$=4 (circles),
  $\langle s \rangle$=6 (squares) and $\langle s \rangle$=10 (triangles).
  For $\sigma \to 0$
  (full symbols) the Gaussian distribution becomes a $\delta$ function with fixed size
  of dependency clusters and thus $p_c$ is identical to those obtained
  by Eqs. (\ref{critical_u_fixed}) and (\ref{p_c}) for groups of single size, $s$.}
\label{PcVsSigma}
\end{figure}

\section{Gaussian Distribution of Dependency Groups}
Using the general solution, given in Eqs. (\ref{u_trans}) and (\ref{P_inf_u}), one can calculate
$P_\infty$ after initial removal of $1-p$ of the nodes and get $p_c$ for every distribution of sizes of
dependency clusters, $q(s)$. Here, we calculate $p_c$ in the case of a normal Gaussian distribution
for the size of the dependency clusters with average size
$\langle s\rangle$, and variance $\sigma^2$. In this case, the probability of a random node to belong to a dependency
  cluster of size $s$ is given by

\begin{eqnarray}
q(s)=\left\{
\begin{tabular}{ccc}
  $Ae^{-(s-\langle s\rangle)^2/2\sigma^2} $ &  & $1<s< 2\langle s\rangle-1$\\
  $0$ &  & elsewhere, \\
\end{tabular}
\right.
\end{eqnarray}

\noindent where $A$ is a normalization constant. Note that $q(s)\neq0$ only for
$1<s< 2\langle s\rangle-1$ in order to have a symmetrical distribution
around $\langle s\rangle$.

  This case generalizes our results of dependency
clusters (for $\sigma\rightarrow 0$) having a single size $s$, and show the deviations from these results as the distribution becomes
broader. In this case there are nodes that belong to dependency clusters which are larger then $\langle s\rangle$ and thus
have higher probability to become non-functional while the same number of nodes belong to dependency clusters which are smaller
 then $\langle s\rangle$ and have smaller probability to become non-functional. We find that the first order transition threshold
 decreases as the distribution of dependency groups becomes broader and thus, the network is more stable, as shown in Fig. \ref{PcVsSigma}(a).
 However, this effect becomes weaker for larger $\langle s\rangle$,
as shown in Fig. \ref{PcVsSigma}(b).


\section{Poisson Distribution of Dependency Groups}
Next, we study the case of a Poisson distribution of dependency cluster sizes.  In this case, the
probability that a random node is depended on $s'$ other nodes (and thus the size of the cluster, $s$,
equal to $s'+1$) is
\begin{equation}\label{p(k)}
    p(s')=\frac{\lambda^{s'} e^{-\lambda}}{{s'}!}\equiv \lim_{n\to\infty}  {n \choose {s'}}
    \left( \frac{\lambda}{n}\right)^{s'} \left( 1-\frac{\lambda}{n}  \right)^{n-{s'}}
\end{equation}
when $\lambda \equiv \langle s\rangle -1$ is the average number of other nodes that depend on a random node.
The dependency process can be calculated, using Eq. (\ref{gD}) and considering that $q(s)=p(s-1)$,
\begin{equation}\label{}
    g_D(T)=\sum_{s=1}^{\infty}p(s-1)T^{s-1}=\sum_{s'=0}^\infty p(s')T^{s'},
\end{equation}
with $s'\equiv s-1$. Using (\ref{p(k)}) we obtain,

\begin{eqnarray*}
  \nonumber g_D(T)&=& \sum_{s'=0}^\infty \lim_{n\to\infty}  {n \choose s'}
    \left( \frac{\lambda}{n}\right)^{s'} \left( 1-\frac{\lambda}{n}  \right)^{n-s'} T^{s'}\\
  \nonumber &=&  \sum_{s'=0}^\infty   {n \choose s'}
    \left( \frac{\lambda}{n}T \right)^{s'} \left( 1-\frac{\lambda}{n}  \right)^{n-s'} \\
             &=& \lim_{n\to\infty} \left[ \frac{\lambda(T-1)}{n}+1 \right]^n \\
             &=& e^{\lambda(T-1)}=e^{(\langle s\rangle -1)(T-1)}.
\end{eqnarray*}

Following (\ref{x_eq}) and (\ref{trans_eq_u}), we get an equation for $u$ for the case of Poisson
distribution:

\begin{equation}\label{u_trans_Bino}
    \ln u=-kp(1-u)e^{(\langle s\rangle -1)(p(1-u)-1)}.
\end{equation}
Finally, using (\ref{P_inf_ln_u}) $P_\infty$ is obtained in a closed form,
\begin{equation}\label{P_inf_trans_Bino}
P_\infty=f_{k,p}(P_\infty)e^{(\langle s\rangle -1)(pf_{k,p}(P_\infty)-1)}
\end{equation}
where $f_{k,p}(P_\infty)\equiv1-\exp{(-kpP_\infty)}$.

\begin{figure}
\begin{center}
\epsfig{file=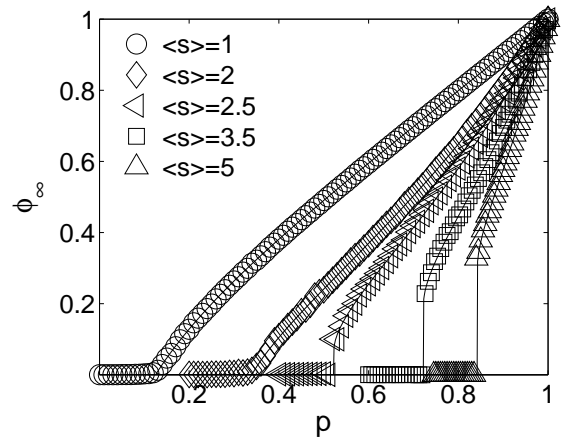,width=8cm}
\end{center}
\caption{The size of the giant cluster, $\phi_\infty\equiv P_\infty \cdot p$, vs.
$p$, for ER networks ($k=8$)
 and Poisson distribution of dependency clusters with different average sizes, $\langle s\rangle$.
The symbols represent simulation results of systems of $50,000$ nodes and the solid lines
show the theoretical predictions. For $\langle s\rangle=1$ and $2$ the network undergoes a second
order transition while for $\langle s\rangle=2.5, 3.5$ and $5$ the network undergoes a first
order transition (see figure (\ref{PcVssPoisson_transition_pI_pII}) where the exact
transition point from first to second order transition is shown).}
\label{PinfVsP_PoissonSize}
\end{figure}

\begin{figure}[!t]
\begin{center}
\begin{tabular}{cc}
\epsfig{file=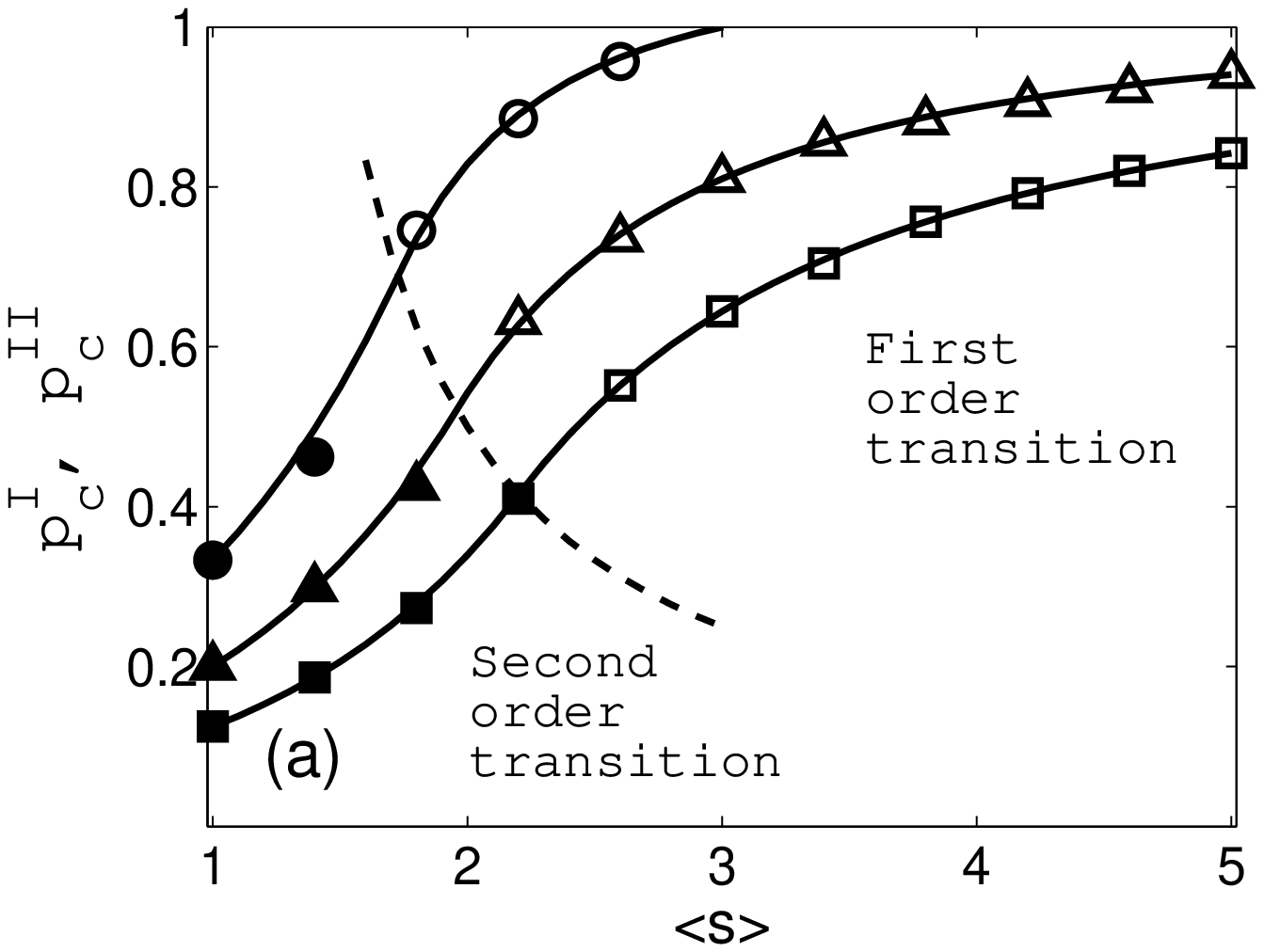,width=8cm}\\
\epsfig{file=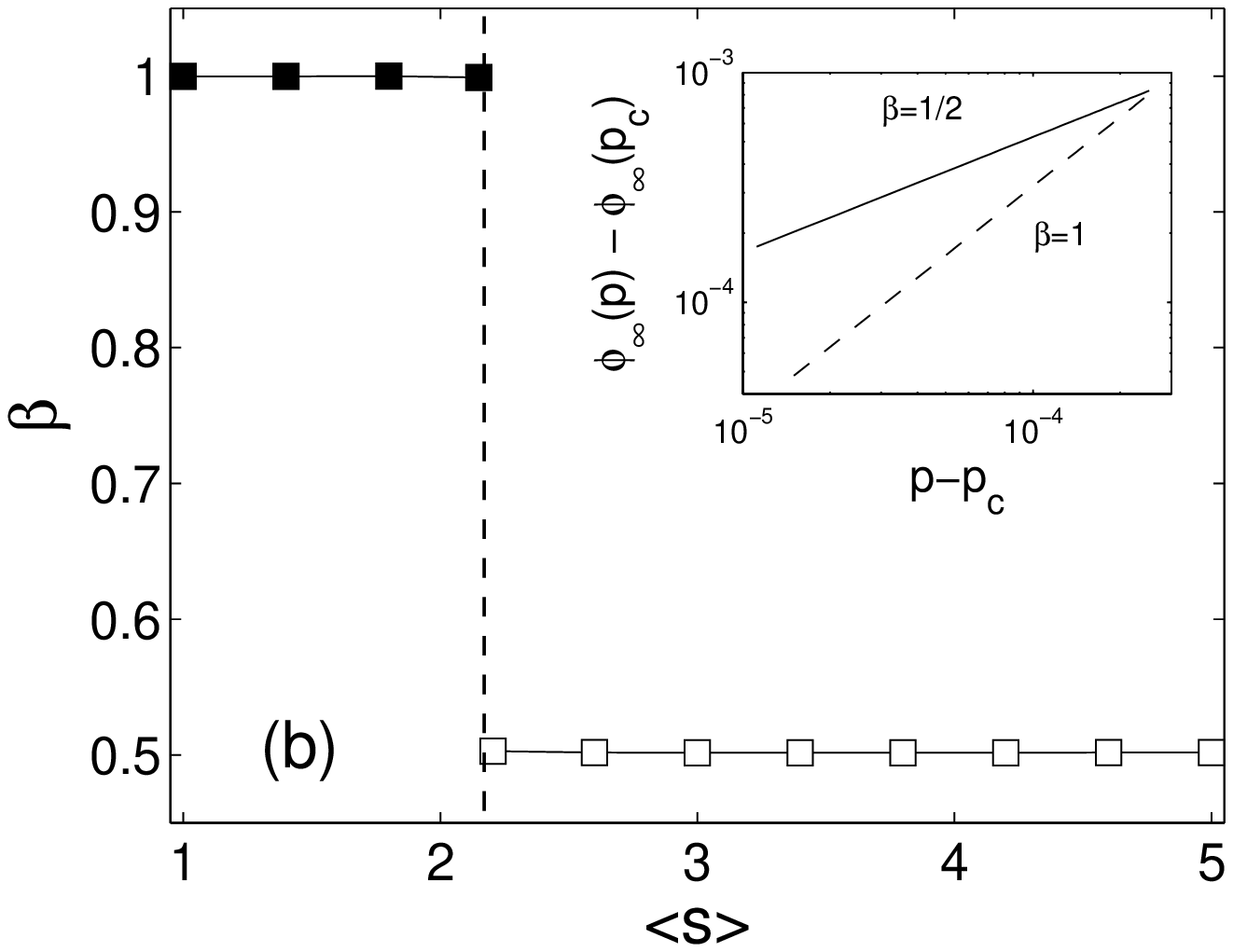,width=8cm}
\end{tabular}
\end{center}
\caption{(a) Theory (lines) and simulations (symbols) are compared for the values
of $p_c^I(\langle s\rangle)$ and $p_c^{II} (\langle s\rangle)$ for ER networks with different
average degree, k (circles for $k=3$, triangles for $k=5$ and squares for $k=8$).
For $\langle s\rangle > \langle s\rangle_c$ - (dashed line)
the network undergoes a first order transition. The theoretical values of
the transition point, $p_c^I(\langle s\rangle)$, that are calculated according to
Eq. (\ref{derivatives_Bino}) are compared with simulations (open symbols) performed using the NOI method
(explained in text). For $\langle s\rangle<\langle s\rangle_c$
the network undergoes a second order transition, and the theoretical values of
the transition point, $p_c^{II} (\langle s\rangle)$, that are calculated according to
Eq. (\ref{P_II}) are compared with simulations (full symbols) performed using the second largest
cluster method. The dashed line separating between the first and
second order is obtained according to Eq. (\ref{transition_pI_pII_pc_s}).
(b) The size of the giant cluster, $\phi_\infty$ above the critical point, $p_c$, is described by
$\phi_{\infty}(p)-\phi_{\infty}(p_c)\sim(p-p_c)^{\beta}$, as shown in the inset.
Critical exponent, $\beta$, above the transition point, $p_c$, is plotted versus the average size of the dependency
clusters, $\langle s\rangle$, for a network with $k=8$ (compare to panel (a); squares).
For the region of the second order transition, $\langle s\rangle<\langle s\rangle_c$, we find $\beta=1$
($\langle s\rangle_c$ is marked by the vertical dashed line) while for the region of the first order transition,
$\langle s\rangle>\langle s\rangle_c$,  $\beta=1/2$. }

\label{PcVssPoisson_transition_pI_pII}
\end{figure}

Fig. \ref{PinfVsP_PoissonSize} shows the size of the giant cluster at steady state versus
$p$ for different values of $\langle s\rangle$. For small $\langle s\rangle$, many nodes do
 not depend on other nodes so the effect
of the dependency clusters is rather weak and thus the percolation transition
is of second order, meaning that $P_\infty$ continuously decreases from a finite value to zero
at a specific transition point, $p_c^{II}$.
However, for large $\langle s\rangle$ the network undergoes a first order transition, meaning
that the size of the giant cluster abruptly jumps discontinuously from finite size
for $p>p_c^I$ to zero for $p<p_c^I$. Such a network is
qualitatively more vulnerable than a network that undergoes a second order transition
due to the cascading failures leading to high vulnerability of the network around $p_c^I$.

Next, we find explicitly, by analyzing Eq. (\ref{u_trans_Bino}), the first order transition point,
 $p_c^I$, and the second order transition point, $p_c^{II}$. Eq. (\ref{u_trans_Bino}) has a trivial solution
for $u=1$, which means that the network is completely fragmented.
The second order transition point, $p_c^{II}$, corresponds to the solution of Eq. (\ref{u_trans_Bino})
where $u \to 1$. This condition gives $p_c^{II}$,
\begin{equation}\label{P_II}
    p_c^{II}=\frac{e^{\langle s\rangle-1}}{k}.
\end{equation}
Note, that for the case of $\langle s\rangle=1$, meaning that all the nodes are not depended,
$p_c^{II}=1/k$ as for regular Erd$\ddot{o}$s-R$\acute{e}$nyi network.

The first order transition point, $p_c^I$, corresponds to the tangential intersection of the left and
right terms of Eq. (\ref{u_trans_Bino}), meaning that the derivatives of both with respect
to $u$ are equal. This yields,

\begin{equation}\label{derivatives_Bino}
     p_c^I(\langle s\rangle-1)=\frac{1}{u-1}-\frac{1}{u\ln u},
\end{equation}
where u is the solution of Eq. (\ref{u_trans_Bino}).

Fig. \ref{PcVssPoisson_transition_pI_pII} (a) shows $p_c^I$ and $p_c^{II}$ versus
the average size of the dependency clusters, $\langle s\rangle$.
At critical values  $p=p^\ast_c$ and $\langle s\rangle=\langle s\rangle_c$
the phase transition changes from first order to a second order.
The values of $p^\ast_c$ and $\langle s\rangle_c$ are obtained when the conditions for both
the first and second order transitions are satisfied simultaneously.
Applying both conditions we obtain

\begin{eqnarray}
\label{transition_pI_pII_k_s}
  2(\langle s\rangle_c -1) &=&  ke^{-(\langle s\rangle_c -1)} , \\
  k &=& \frac{1}{p^\ast_c}e^{1/2p^\ast_c} . \label{transition_pI_pII_k_pc}
\end{eqnarray}

For a given ER network with average degree $k$, Eq. (\ref{transition_pI_pII_k_s})
provides the critical average size of dependency clusters, $\langle s\rangle_c$.
A network with  $\langle s\rangle < \langle s\rangle_c$ undergoes a second order
phase transition while for $\langle s\rangle > \langle s\rangle_c$ the network
undergoes a first order transition. Therefore, $p^\ast_c$, obtained from Eq.
(\ref{transition_pI_pII_k_pc}) for the case of $\langle s\rangle = \langle s\rangle_c$,
characterizes the stability of a network with maximal $\langle s\rangle$
under the constraint of undergoing second order transition.
The critical case, shown in Fig. \ref{PcVssPoisson_transition_pI_pII} (a),
of transition from first order to second order transition (dashed line) is given by

\begin{equation}\label{transition_pI_pII_pc_s}
    p^\ast_c = \frac{1}{2(\langle s\rangle_c -1)}.
\end{equation}

As shown in Figs. \ref{PinfVsP_PoissonSize} and \ref{PcVssPoisson_transition_pI_pII}, increasing the size of
the dependency clusters increases the network vulnerability and $p_c^I$ becomes larger.
A critical average size of dependency clusters, $\langle s\rangle_{max}$, corresponds to
$p_c^I=1$ meaning that the network completely fragments as a result of removal of even a single node.
Such a network can be regarded as unstable. The value of $\langle s\rangle_{max}$ is given by Eqs.
(\ref{derivatives_Bino}) and (\ref{u_trans_Bino}) under the condition of $p_c^I=1$.

Thus, the stability of a random network with dependency clusters with average size
$\langle s\rangle$ with Poissonian distribution can be summarized
\begin{tabbing}
  Size of dependency cluster \= Stability \kill
$\langle s\rangle < \langle s\rangle_c$ \> Second order transition \\
 $\langle s\rangle_c < \langle s\rangle < \langle s\rangle_{max}$ \> First order transition \\
$  \langle s\rangle \geq \langle s\rangle_{max}$ \> Unstable network.
\end{tabbing}

We thank Avi Gozolchiani for helpful discussions. We thank the European EPIWORK project, the Israel Science Foundation,
ONR, DFG, and DTRA for financial support.


\end{document}